\documentclass[a4paper,11pt]{article}

\usepackage{jcappub}
\usepackage[latin1]{inputenc}
\usepackage{amsmath}
\usepackage{hyperref}
\hypersetup{colorlinks=True,citecolor=blue}
\usepackage{amsfonts}
\usepackage{amssymb}
\usepackage{caption}
\usepackage{subfigure}
\usepackage{float}
\usepackage{bm}
\usepackage{graphicx}
\usepackage{color}
\usepackage{verbatim}

\graphicspath{{}}

\def\Eq#1{{Eq.~(\ref{#1})}}

\title{Constraining stochastic gravitational wave 
background from weak lensing of CMB  B-modes}

\author[\dagger]{Shabbir Shaikh,} 
\emailAdd{shabbir@iucaa.in} 
\affiliation [\dagger]{Inter University Centre for Astronomy and Astrophysics, Post Bag 4, Ganeshkhind, Pune-411007, India}

\author[\dagger]{Suvodip Mukherjee,}
\emailAdd{suvodip@iucaa.in}

\author[\ddagger]{Aditya Rotti,}
\emailAdd{adityarotti@gmail.com}

\affiliation [\ddagger]{Department of Physics, Florida State University, Tallahassee, FL 32304, USA}

\author[\dagger]{and Tarun Souradeep} 
\emailAdd{tarun@iucaa.in}

\date{\today}

\abstract{A stochastic gravitational wave background (SGWB) will affect the CMB anisotropies via weak lensing.
Unlike weak lensing due to large scale structure which only deflects photon trajectories, a SGWB has an additional effect of rotating the polarization vector along the trajectory. We study the relative importance of these two effects, deflection \& rotation, specifically in the context of E-mode to B-mode power transfer caused by weak lensing due to SGWB. Using weak lensing distortion of the CMB as a probe, we derive constraints on the spectral energy density ($\Omega_{GW}$) of the SGWB, sourced at different redshifts, 
without assuming any particular model for its origin. We present these bounds on $\Omega_{GW}$ for different power-law models characterizing the SGWB, indicating the threshold above which observable imprints of SGWB must be present in CMB.}

\begin{document}

\maketitle

\section{Introduction}

\par The Cosmic Microwave Background (CMB) is an exquisite tool to study 
the universe. It is being used to probe the early universe scenarios as well as 
the physics of processes happening in between the surface of last scattering and 
the observer. Well studied processes among these include lensing by large scale structure, Sunyaev-Zeldovich effect, 
integrated Sachs-Wolfe effect etc. These effects give rise to secondary anisotropies in the CMB. The stochastic  gravitational 
wave background (SGWB), if present, will affect the CMB via weak lensing \cite{Mollerach_cmb_gw_lensing, li_and_cooray}. 
The SGWB can be sourced by inflation, astrophysical phenomena like halo mergers and halo formation 
\cite{Takahiro_SGWB_halo_mergers, Carbone:2005nm},  second order density perturbations 
\cite{devdeep_s_cosmic_shear}, early universe phase transitions \cite{Maggiore2000283}, etc.
In the new era, post the first direct detection of gravitational wave by LIGO \cite{PhysRevLett.116.061102} and studies 
assessing a SGWB for such populations \cite{PhysRevLett.116.131102}, a reassessment of SGWB probed by weak lensing of CMB considered 
earlier \cite{aditya_rotti_new_window_into_SGWB} appears to be timely. 

Effects of lensing by scalar and tensor perturbations on CMB have been calculated in full detail in 
literature \cite{li_and_cooray, 
cooray_kamionkowdki_caldwell, dodelson_rozo_stebbins, hamsa_padmanabhan_lensing_kernel, liang_dai_rotation}.
Padmanabhan et al. (\cite{hamsa_padmanabhan_lensing_kernel}) carried out a comparative study of 
lensing by scalar and tensor perturbations, concluding that  tensor perturbations are more efficient than scalar perturbations at converting E-modes of CMB polarization
 to B-modes. More recently, Dai \cite{liang_dai_rotation} noted the effect of the rotation of  CMB 
polarization due to tensor perturbations, arguing that the B-mode power generated by lensing 
deflection due to tensor perturbations is largely canceled by the rotation of polarization induced 
by these perturbations. 
In summary, unlike in the case of weak lensing by large scale structure, a SGWB leads to 
two different effects in CMB: (i) deflection of photon 
path and (ii) rotation of polarization vector of photon along the direction of propagation. 
The SGWB results in additional distortions in the CMB sky, over and above those introduced by lensing due to large scale structure. 

It has been shown that the lensing due to SGWB sourced by inflation is below the cosmic variance and 
hence not detectable even for cosmic variance limited experiments \cite{li_and_cooray}. However, in 
light of other conjectured sources of SGWB, 
weak lensing of CMB by SGWB has been used in previous work \cite{aditya_rotti_new_window_into_SGWB} 
to derive upper bounds on $\Omega_{GW}$. Namikawa et al. \cite{Namikawa:2014lla} have studied the detectability of 
weak lensing of CMB induced by gravitational waves. They do not include the effect of the rotation of CMB 
polarization in their evaluations.

In this paper, we carry out a more careful assessment of the efficiency of tensor perturbations in 
mediating power transfer between E-mode and B-mode of CMB polarization. Finally, we incorporate 
rotation effect in the lensing kernels and derive revised constraints on the energy density 
$\Omega_{GW}$ of the SGWB, for different empirical models 
of SGWB power generated at a number of representative source redshifts.

This paper is organized as follows. In section II we carefully assess the relative contributions of rotation and deflection associated with weak lensing due to the SGWB. 
In section III we present the details of the procedure used to derive the revised upper limits on $\Omega_{GW}$. We conclude with the discussion of our results in section IV.
We use the best fit Planck+WP+highL+BAO parameters from Planck 2013 \cite{Ade:2013zuv}  to derive 
all our results.

 \section{Weak lensing of CMB by gravitational waves}
\par Weak lensing of CMB remaps the temperature and polarization anisotropy 
field on the sky. The lensed temperature anisotropy $\tilde T(\hat n)$, observed in the direction 
$\hat n$ corresponds to the temperature anisotropy $T(\hat n + \vec{d})$,
observed in the absence of lensing in the direction $\hat n + \vec{d}$,  
\begin{equation}\label{eq_T}
 \tilde T(\hat n) = T(\hat n + \vec{d}),
\end{equation}
where $\vec{d}$ is the deflection angle and defines a vector field on the sky. CMB photons are linearly polarized 
because of Thomson scattering. CMB polarization field is expressed using $Q$ and 
$U$ Stokes parameters, ${}_{\pm}X(\hat n) = Q(\hat{n})\pm iU(\hat{n})$.  
To consider the complete effect of 
lensing on polarization anisotropies, we have to consider the rotation 
of polarization vector of CMB photons about its direction of propagation due to metric 
perturbations as described by Dai \cite{liang_dai_rotation}. Including the effect 
of photon deflection and rotation of polarization, the lensed polarization field is 
described as:
\begin{equation}\label{eq_tildeX}
{}_{\pm} \tilde X(\hat n) = e^{\mp 2i \psi (\hat n)}  {}_{\pm}X(\hat n + \vec{d}),
\end{equation}
where $\psi$ is the angle of rotation of polarization.

The vector deflection angle is field decomposed into a gradient potential $\phi (\hat n)$ and 
a curl potential $\Omega(\hat n)$
\begin{equation} \label{eq_alpha}
d_{i} = \nabla_{i} \phi(\hat n) - \varepsilon_{i}^{jk} n_{j} \nabla_{k} \Omega(\hat n),
\end{equation}
where $\vec{\nabla}$ is the angular gradient on the sphere. For a statistically
isotropic lensing field, $\phi (\hat n)$ and $\Omega (\hat n)$ are described by 
their angular power spectrum $C_{l}^{\phi \phi}$ and $C_{l}^{\Omega \Omega}$ 
respectively. Methods to reconstruct both $C_{l}^{\phi \phi}$ and $C_{l}^{\Omega \Omega}$ from 
the observed CMB sky exist in literature, for example, see \cite{Okamoto:2003zw, Namikawa:2011cs} 
and references therein.
The rotation angle $\psi(\hat n)$ is related to curl potential $\Omega (\hat n)$ through 
\cite{liang_dai_rotation}
\begin{equation}\label{eq_psi}
 \psi (\hat{n}) = -\frac{1}{2} \nabla^{2}\Omega(\hat{n}),
\end{equation}
where $\nabla^{2}$ is angular Laplacian. \Eq{eq_psi} shows that the source of 
the curl potential $\Omega (\hat n)$ gives rise to the rotation of polarization vector. 
Angular power spectrum for rotation $C_{l}^{\psi \psi}$ is related to $C_{l}^{\Omega \Omega}$
through \cite{liang_dai_rotation}
\begin{equation}\label{eq_clpsipsi}
 C_{l}^{\psi \psi} = [l(l+1)/2]^{2}C_{l}^{\Omega \Omega},
\end{equation}
and the deflection-rotation cross power spectrum is 
\begin{equation}\label{eq_clpsiomega}
 C_{l}^{\psi \Omega} = [l(l+1)/2]C_{l}^{\Omega \Omega}.
\end{equation}
Note that $C_{l}^{\psi \psi}$ is $\sim l^{2} C_{l}^{\psi \Omega}$, which makes 
$C_{l}^{\psi \psi}$ much stronger over $C_{l}^{\psi \Omega}$ at small angular 
scales (high multipoles $l$).

At the linear order in perturbation, lensing by large scale structure (LSS) in the 
universe, which corresponds to scalar metric perturbations, induce only
gradient type deflections. Gravitational waves, which corresponds to tensor
metric perturbations, induce both gradient and curl type deflections even at
linear order \cite{Stebbins:1996wx, dodelson_rozo_stebbins, cooray_kamionkowdki_caldwell}.
Hence, to consider the complete effect of curl deflection sourced by 
scalar and tensor perturbations at linear order, 
we include the rotation of polarization in our computation. 
However, we neglect the scalar deflection caused by the tensor perturbations, because 
it is an order of magnitude less than the tensor deflection \cite{li_and_cooray}. 
There are several models predicting vector perturbations which can also 
contribute to curl deflections, for example, vector perturbations caused by cosmic strings \cite{Yamauchi:2013fra}. 
Since the relative amplitude and spectrum of vector perturbations would be model dependent,
we choose to neglect the lensing by vector perturbations in our analysis.

Effect of lensing on CMB angular power spectrum is computed either using
real space correlation function \cite{PhysRevD.71.103010} or using spherical 
harmonic space correlation function method \cite{PhysRevD.62.043007}. Here we 
have provided the expressions obtained using latter method, originally computed
in \cite{PhysRevD.62.043007} for scalar deflection, in \cite{li_and_cooray} and 
\cite{hamsa_padmanabhan_lensing_kernel} for scalar and tensor deflection and in  
\cite{liang_dai_rotation} for scalar and tensor deflection including the 
effect of rotation.

Lensed TT angular power spectrum is:
\begin{align} \label{eq_lClTT}
\begin{split}
 \tilde{C}_{l}^{TT} = C_{l}^{TT} - l(l+1) R C_{l}^{TT} +  
 \frac{1}{2l+1} \sum_{l_{1}l_{2}} C_{l_{2}}^{TT} [(F_{ll_{1}l_{2}}^{\phi})^2 C_{l_{1}}^{\phi}  
P^{+}_{ll_{1}l_{2}}
+(F_{ll_{1}l_{2}}^{\Omega})^2 C_{l_{1}}^{\Omega} P^{-}_{ll_{1}l_{2}}],
\end{split}
\end{align}
where $R$ is the rms deflection power given by
\begin{eqnarray} \label{eq_R}
 R = \sum_{l} \frac{l(l+1)(2l+1)}{8\pi} (C_{l}^{\phi \phi} + C_{l}^{\Omega 
\Omega}).
\end{eqnarray}
$R$ is measure of rms deflection angle, $d^{2}_{rms} = R$. 
$F_{ll_{1}l_{2}}^{\phi}$ and $F_{ll_{1}l_{2}}^{\Omega}$ are lensing kernels:

\begin{eqnarray} \label{F_phi}
F_{ll_{1}l_{2}}^{\Omega} = F_{ll_{1}l_{2}}^{\phi} = 
-\sqrt{l_{1}(l_{1}+1)l_{2}(l_{2}+1)} 
\sqrt{\frac{\Pi_{l l_1 l_2}}{4\pi}}
\Big( \begin{array}{lrc} l & l_{1} & l_{2} \\
      0 & -1 & 1 \end{array} \Big), 
\end{eqnarray}
where $\Pi_{l l_1 ...} = (2l+1)(2l_1 +1)...$ and $P^{\pm}_{ll_{1}l_{2}} = (1 \pm 
(-1)^{l+l_{1}+l_{2}})/2$.
It is clear from \Eq{eq_lClTT} that rotation of 
polarization has no contribution in the lensing of temperature anisotropy.

\par
\begin{figure}[H]
    \subfigure[Bin-1, $l' = 2-50$]{\label{fig1a}  
\includegraphics[height=6cm,width=8cm,angle=0]{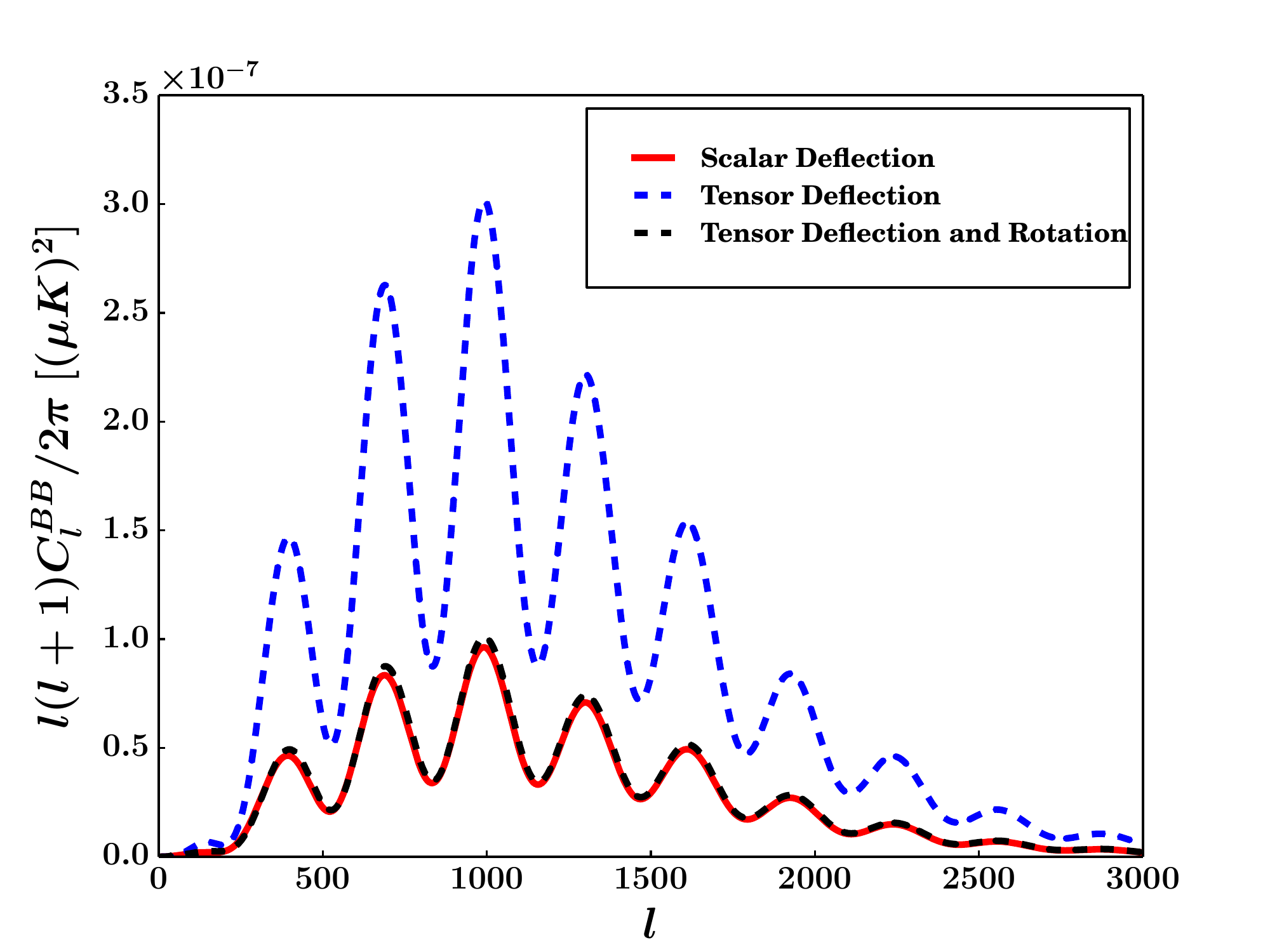}}
    \subfigure[Bin-2, $l' = 51-100$]{\label{fig1b}  
\includegraphics[height=6cm,width=8cm,angle=0]{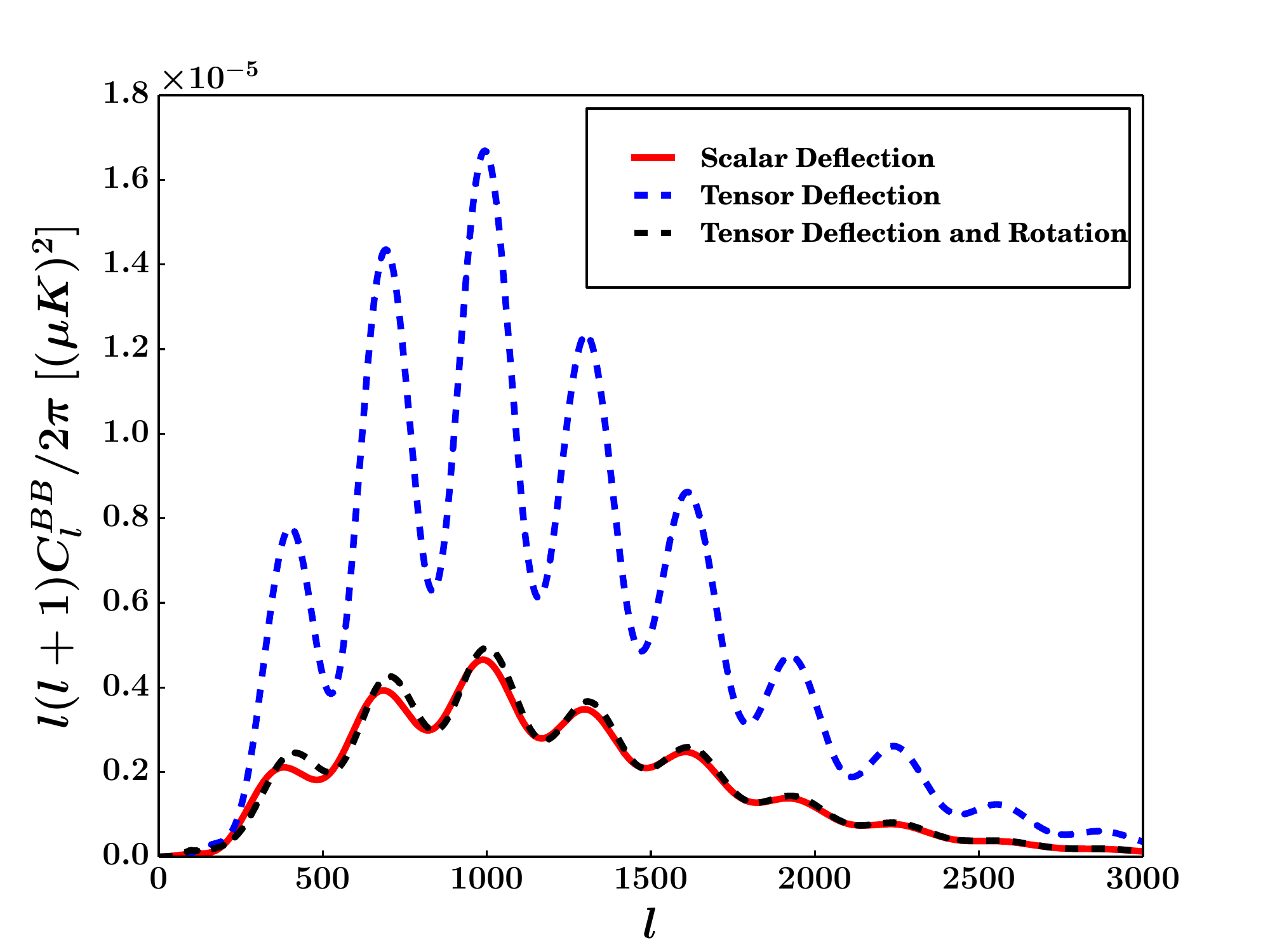}}
    \subfigure[Bin-3, $l' = 201-250$]{\label{fig1c}  
\includegraphics[height=6cm,width=8cm,angle=0]{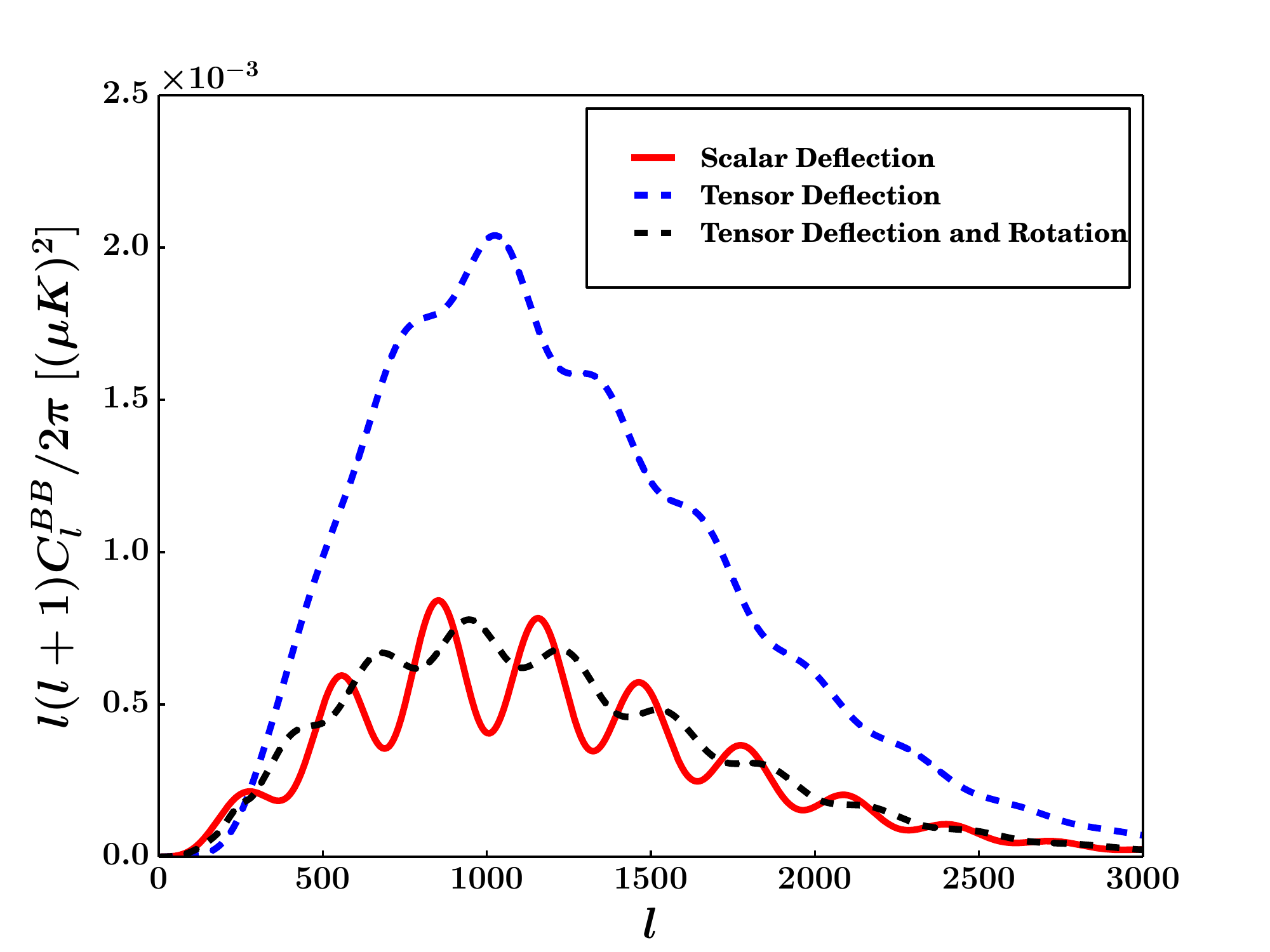}}
    \subfigure[Bin-4, $l' = 451-500$]{\label{fig1d}  
\includegraphics[height=6cm,width=8cm,angle=0]{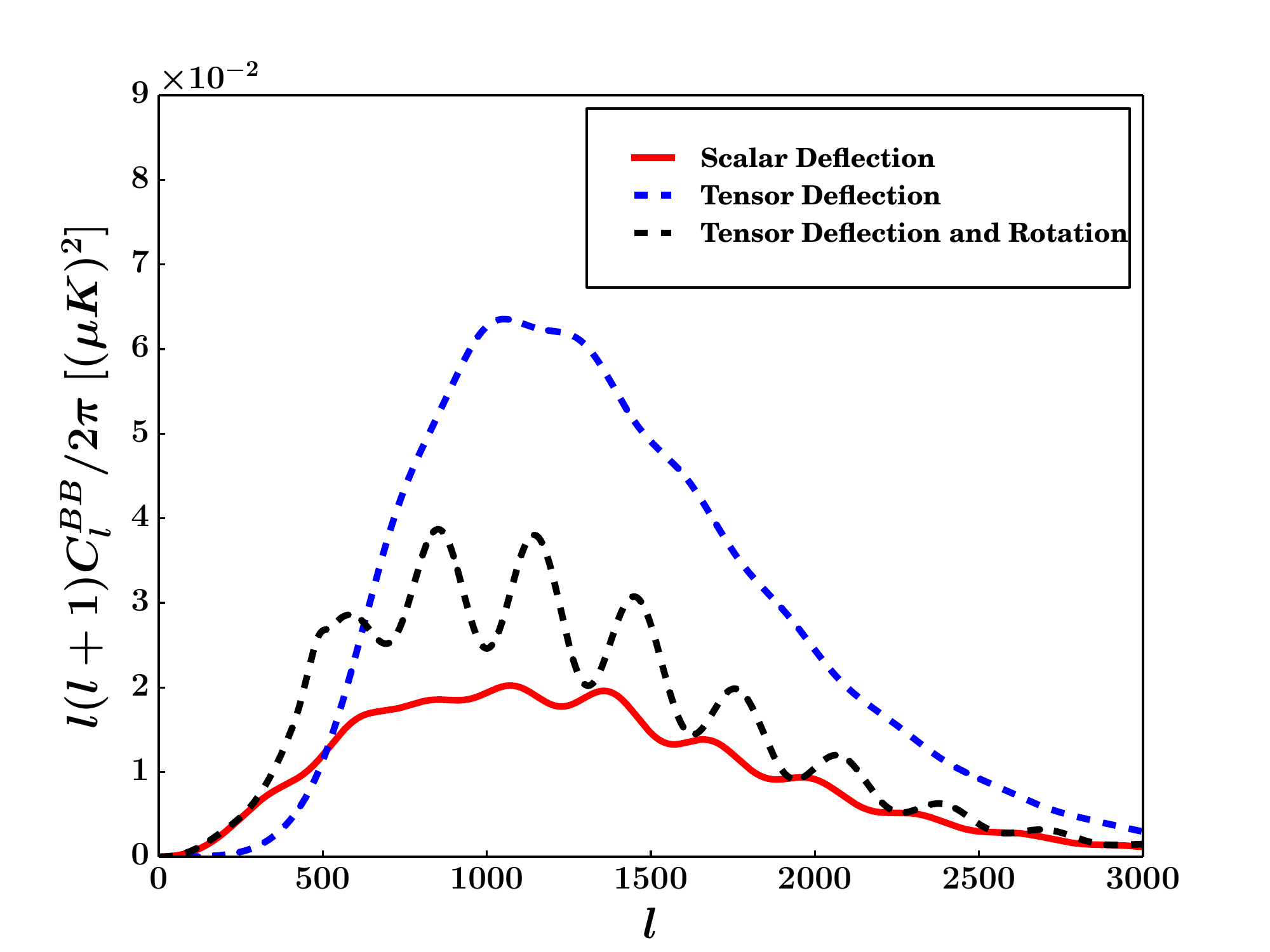}}
    \subfigure[Bin-5, $l' = 951-1000$]{\label{fig1e}  
\centerline{\includegraphics[height=6cm,width=8cm,angle=0]{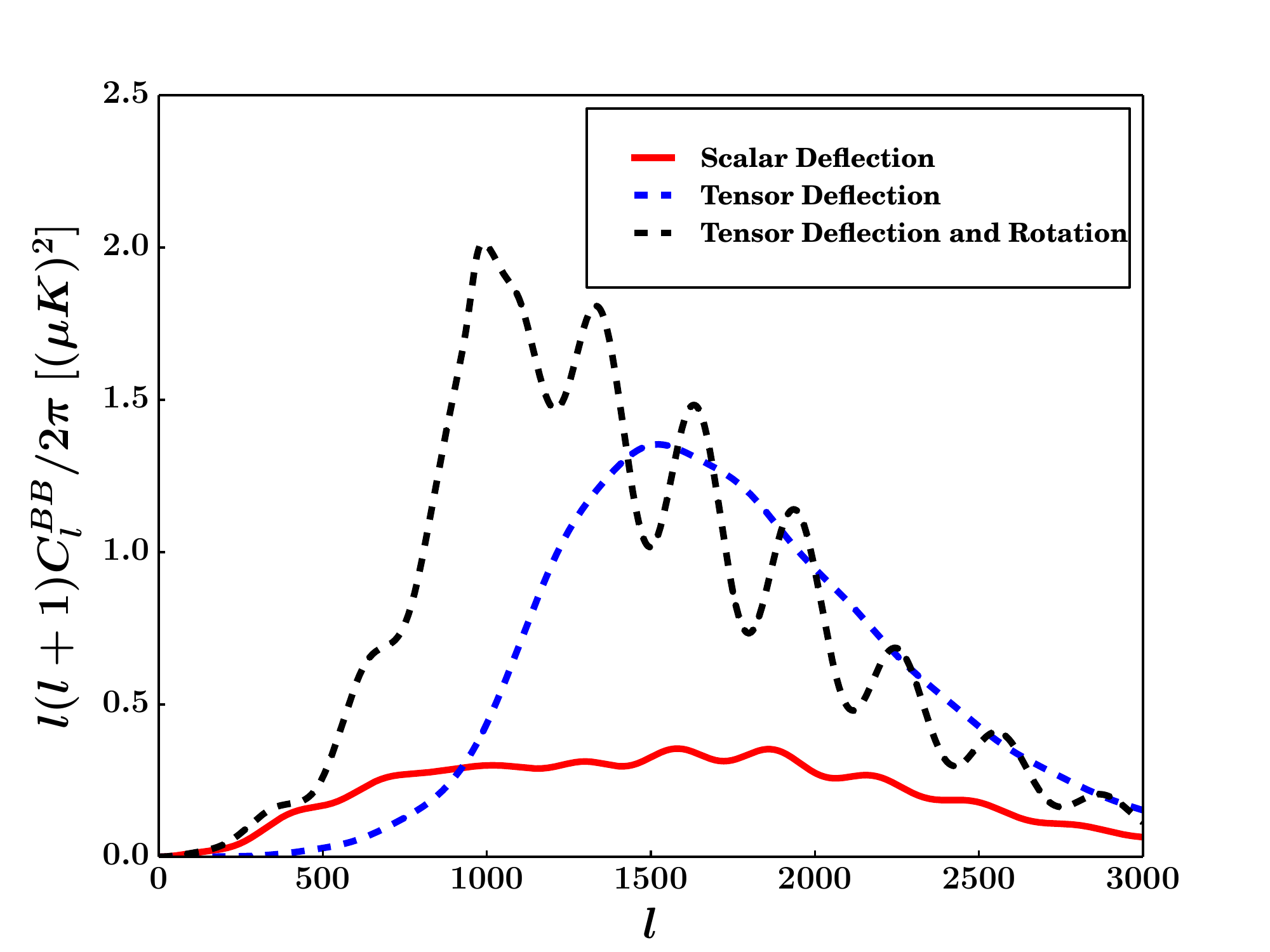}}}
		\captionsetup{singlelinecheck=off,justification=raggedright}
	 \caption{Comparison of individual contribution to lensed B-mode power spectrum due to
	 scalar deflection, tensor deflection and tensor deflection + rotation for
	 lensing power in five different bins. $l'$ denotes the bin range of multipole 
	 over which $C_{l'}^{\Omega \Omega}$ has nonzero power = $10^{-17}$}
	  \label{fig1}
\end{figure}

 Polarization E mode and B mode angular power spectra are 
\begin{eqnarray}\label{eq_lClEE} 
\tilde{C}_{l}^{EE} &=& C_{l}^{EE} - (l^{2}+l-4) R C_{l}^{EE} - 4 S C_{l}^{EE}
\nonumber \\
&+& \frac{1}{(2l+1)} \sum_{l_{1}l_{2}} [C_{l_{1}}^{\phi \phi} 
({}_{2}F_{ll_{1}l_{2}}^{\phi})^2 (C_{l_{2}}^{EE} P^{+}_{ll_{1}l_{2}}+ C_{l_{2}}^{BB} P^{-}_{ll_{1}l_{2}}) 
+ C_{l_{1}}^{\Omega \Omega} ({}_{2}F_{ll_{1}l_{2}}^{\Omega})^2 
(C_{l_{2}}^{EE} P^{-}_{ll_{1}l_{2}}+ C_{l_{2}}^{BB} P^{+}_{ll_{1}l_{2}})]\nonumber \\
&+& \frac{4}{(2l+1)} \sum_{l_{1}l_{2}} [C_{l_{1}}^{\psi \psi} 
({}_{2}F_{ll_{1}l_{2}}^{\psi})^2 - C_{l_{1}}^{\Omega \psi} 
{}_{2}F_{ll_{1}l_{2}}^{\Omega} {}_{2}F_{ll_{1}l_{2}}^{\psi}] 
(C_{l_{2}}^{EE} P^{-}_{ll_{1}l_{2}}+ C_{l_{2}}^{BB}P^{+}_{ll_{1}l_{2}}),
\end{eqnarray}

\begin{eqnarray}\label{eq_lClBB} 
\tilde{C}_{l}^{BB} &=& C_{l}^{BB} - (l^{2}+l-4) R C_{l}^{BB} - 4 S C_{l}^{BB}
\nonumber \\
&+& \frac{1}{(2l+1)} \sum_{l_{1}l_{2}} [C_{l_{1}}^{\phi \phi} 
({}_{2}F_{ll_{1}l_{2}}^{\phi})^2 (C_{l_{2}}^{EE} P^{-}_{ll_{1}l_{2}}+ C_{l_{2}}^{BB} P^{+}_{ll_{1}l_{2}}) 
+ C_{l_{1}}^{\Omega \Omega} ({}_{2}F_{ll_{1}l_{2}}^{\Omega})^2 
(C_{l_{2}}^{EE} P^{+}_{ll_{1}l_{2}}+ C_{l_{2}}^{BB} P^{-}_{ll_{1}l_{2}})]\nonumber \\
&+& \frac{4}{(2l+1)} \sum_{l_{1}l_{2}} [C_{l_{1}}^{\psi \psi} 
({}_{2}F_{ll_{1}l_{2}}^{\psi})^2 - C_{l_{1}}^{\Omega \psi} 
{}_{2}F_{ll_{1}l_{2}}^{\Omega} {}_{2}F_{ll_{1}l_{2}}^{\psi}] 
(C_{l_{2}}^{EE} P^{+}_{ll_{1}l_{2}}+ C_{l_{2}}^{BB}P^{-}_{ll_{1}l_{2}}),
\end{eqnarray}
$S$ is the rms rotation power given by
\begin{equation} \label{S}
 S = \sum_{l} \frac{2l+1}{4\pi} C_{l}^{\psi \psi}.
\end{equation}
${}_{2}F_{ll_{1}l_{2}}^{\phi}$ and ${}_{2}F_{ll_{1}l_{2}}^{\Omega}$ are lensing
kernels:
\begin{flalign} \label{2F_phi}
{}_{2}F_{ll_{1}l_{2}}^{\phi/ \Omega} = \sqrt{\frac{l_{1}(l_{1}+1)\Pi_{l l_1 l_2}}{8\pi}} 
\Big[\sqrt{\frac{(l_{2}+2)(l_{2}-1)}{2}} \Big( \begin{array}{lrc} l & l_{1} & l_{2} \\ 
2 & -1 & -1 \end{array} \Big) \pm \sqrt{\frac{(l_{2}-2)(l_{2}+3)}{2}} \Big( 
\begin{array}{lrc} l & l_{1} & l_{2} \\ 2 & 1 & -3 \end{array} \Big) \Big],
\end{flalign}
Here $\Big( \begin{array}{lrc} l & l_{1} & l_{2} \\ m & m_1 & m_2 \end{array} \Big)$ 
denote Wigner-3j symbols.
Lensing kernels ${}_{2}F_{ll_{1}l_{2}}^{\phi}$ and ${}_{2}F_{ll_{1}l_{2}}^{\Omega}$ differ only 
by a negative sign.
${}_{2}F_{ll_{1}l_{2}}^{\psi}$, lensing kernel introduced by rotation, is
\begin{equation} \label{2F_psi}
{}_{2}F_{ll_{1}l_{2}}^{\psi} = \sqrt{\frac{\Pi_{l l_1 l_2}}{4\pi}}
\Big( \begin{array}{lrc} l & l_{1} & l_{2} \\ 2 & 0 & -2 \end{array} \Big).
\end{equation}
TE angular power spectrum is
\begin{eqnarray} \label{eq_lClTE}
 \tilde{C}_{l}^{TE} &=& C_{l}^{TE} - (l^{2}+l-2) R C_{l}^{TE} -2SC_{l}^{TE} 
- \frac{1}{2l+1} \sum_{l_{1}l_{2}} C_{l_{1}}^{\phi \phi} C_{l_{2}}^{TE} 
F_{ll_{1}l_{2}}^{\phi} {}_2 F_{ll_{1}l_{2}}^{\phi} \nonumber \\ 
&-& \frac{1}{2l+1} \sum_{l_{1}l_{2}} C_{l_{1}}^{\Omega \Omega} C_{l_{2}}^{TE} 
F_{ll_{1}l_{2}}^{\Omega} {}_2 F_{ll_{1}l_{2}}^{\Omega} 
+ \frac{1}{2l+1} \sum_{l_{1}l_{2}} C_{l_{1}}^{\Omega \phi} C_{l_{2}}^{TE} 
F_{ll_{1}l_{2}}^{\Omega} {}_2 F_{ll_{1}l_{2}}^{\psi}. 
\end{eqnarray}
To comprehend the effect of both lensing and rotation on $C_l^{BB}$
\footnote{Lensing by tensor perturbations affect 
the $BB$ spectrum more than it affects $TT$, $EE$ and $TE$ spectra \cite{li_and_cooray, 
hamsa_padmanabhan_lensing_kernel}.}, we consider five different 
cases of $C_l^{\Omega\Omega}$ with non-zero constant value $10^{-17}$ over only a  limited $l$ range, mentioned in Fig. \ref{fig1}.
In Fig. \ref{fig1} we plot the individual contribution to lensed $BB$ spectrum due to scalar 
deflection, tensor deflection and tensor 
deflection including rotation. We assume primordial B-modes to be zero. To 
compare the relative contribution of each effect we set $C_{l}^{\phi \phi}  = 
C_{l}^{\Omega \Omega}$. Fig. \ref{fig1} shows, as pointed out in \cite{hamsa_padmanabhan_lensing_kernel},
tensor deflection is more efficient 
than scalar deflection at converting E-mode to B-mode. In the case of Fig. \ref{fig1a}, 
once the contribution of rotation of polarization is 
included, excess B-mode generated by tensor deflection are largely canceled. 
This is in accordance with the results presented by Dai \cite{liang_dai_rotation}.
Dai \cite{liang_dai_rotation} has considered $C_{l}^{\Omega \Omega}$ to be 
caused by tensor perturbations of inflationary origin. $C_{l}^{\Omega \Omega}$ of 
inflationary origin has non-negligible power only up to $l \approx 100$. The case of bin-1 is 
similar to this. Hence Fig. \ref{fig1a} 
verifies the claim of \cite{liang_dai_rotation}. This 
cancellation of excess B-mode is due to correlation between curl 
deflection field and rotation of polarization, $C_{l}^{\psi \Omega}$  given by 
\Eq{eq_clpsiomega}. In the expressions for $\tilde{C}_{l}^{BB}$, term 
containing $C_{l}^{\psi \Omega}$ appears with a negative sign causing the 
cancellation. However we stress that this cancellation is not an exact 
cancellation at each $l$ where excess contribution due to tensor deflection at each 
multipole $l$ is exactly canceled by the contribution due to rotation term 
at that multipole. This depends on the nature of $C_{l}^{\Omega \Omega}$. The 
maximum cancellation of excess B-modes occur when power in 
$C_{l}^{\Omega \Omega}$ is limited to low $l$. In the example shown in Fig. \ref{fig1}, 
maximum cancellation has occurred for bin-1 (Fig. \ref{fig1a}). 
But Fig. \ref{fig1c}, Fig. \ref{fig1d} and Fig. \ref{fig1e} show that the excess 
B-modes by curl deflection are not canceled 
completely once the rotation is included. 
Depending on the nature of $C_{l}^{\Omega \Omega}$ there can be residual power at 
$C_{l}^{BB}$ at large $l$. This is due to the fact that the $C_{l}^{\psi\psi}$ which 
adds with the kernel given in \Eq{eq_lClBB}, is dominant over $C_{l}^{\psi\Omega}$ at 
high $l$. Hence addition due to rotation term becomes important at high $l$. This is evident in 
Fig. \ref{fig1e} where contribution due to tensor deflection with rotation is dominant 
over contribution due to only tensor defection at some values of $l$.
Also, it should be noted that the relative effect of rotation term is most evident when power in 
$C_{l}^{\Omega \Omega}$ is either at 
low $l$ or at high $l$.
\par Lensing potential $C_{l}^{\Omega \Omega}$ induced by SGWB provides a window to constrain 
$\Omega_{GW}$. Different models of generation of tensor perturbations predict different forms 
and amplitudes for $C_{l}^{\Omega \Omega}$ \cite{devdeep_s_cosmic_shear}. Each of 
this lensing potentials may not be detectable on their own. For example, \cite{li_and_cooray}
has shown that lensing potential introduced by inflationary gravitational wave 
background gives the lensing contribution which is below the cosmic variance. 
We do not address any particular model generating the lensing potential. Instead, we assume 
well motivated general forms of lensing potential and assess at what amplitude they produce any detectable effect on CMB 
through lensing. The method used in our analysis is presented in the following section.  

\section{Method}\label{methods}
Curl deflection potential, $C_{l}^{\Omega \Omega}$ is related 
to the energy density of SGWB through the power spectrum of tensor perturbations, 
$P_{H}(k)$. 
Power spectrum of curl deflection potential $C_{l}^{\Omega \Omega}$ is \cite{li_and_cooray}
\begin{equation}\label{eq_clomegaomega}
 C_{l}^{\Omega \Omega} = \frac{\pi}{l^{2}(l+1)^{2}} \frac{(l+2)!}{(l-2)!} \int d^{3}k P_{H}(k)|T_{l}^{H}(k)|^{2},
\end{equation}
where $T_{l}^{H}(k)$ accounts for the evolution of tensor perturbations in the given universe 
and their projection onto the sphere. $T_{l}^{H}(k)$ is given by
\begin{equation}\label{eq_TlH}
 T_{l}^{H}(k) = 2k \int_{\eta_{s}}^{\eta_0} d \eta^{'} T_{H}(k, \eta^{'} - \eta_{s}) \frac{j_{l}(k(\eta_{0} 
 - \eta^{'}))}{(k(\eta_{0} - \eta^{'}))^{2}},
\end{equation}
where $\eta$ is the conformal time. $\eta_{s}$ denotes the conformal time at source 
redshift and $\eta_{0}$ denotes conformal time at present epoch.
$T_{H}(k, \eta)$ is the transfer function for tensor perturbations given by $3 j_{1}(k \eta)/(k 
\eta)$.
$T_{H}$ depends on $(\eta^{'} - \eta_{s})$ and not only on $\eta^{'}$.
We adopt the following definition for the power spectrum $P_{H}(k)$ \cite{li_and_cooray}
\begin{equation}
 \langle H_i(\vec{k}) H^{*}_j(\vec{k^{'}}) \rangle = (2 \pi)^3 P_{H}(k) 
\delta_{ij}\delta^{(3)}(\vec{k} -\vec{k^{'}}),
\end{equation}
where $H(\vec{k})$ is tensor metric perturbation.
Tensor perturbations realized as SGWB contribute to the energy density of the 
universe. Spectral energy density of SGWB ($\rho_{GW}$) at present epoch is generally 
expressed in term of the density parameter $\Omega_{GW}$, which is
\begin{equation}
 \Omega_{GW}(k) = \frac{1}{\rho_{c0}c^{2}} \frac{d\rho_{GW}(k,z = 0)}{d \ln k},
\end{equation}
where $\rho_{c0} = \frac{3 H_0^2}{8 \pi G}$ is the critical density of the universe at the present 
epoch. The spectral energy density, $\Omega_{GW}$ at the present epoch can be expressed as
\begin{equation}\label{eq_Omega_GW}
 \Omega_{GW}(k) = \frac{4 \pi}{3} \Big(\frac{c}{H_{0}} \Big)^{2} k^{3} 
 P_{H}(k) \Big[k \frac{dT_{H}(x)}{dx} \Big]^{2}_{k(\eta_{0} - \eta_{s})}.
\end{equation}
It is known that a power law form of power spectrum $P_{H}(k)$ gives rise to 
the lensing potential $C_{l}^{\Omega \Omega}$ that can be approximated by a power law 
to good accuracy \cite{Laura_Brook_21_cm_lensing_by_GW}. In particular $P_{H}(k) = k^{-n}$ 
gives $C_{l}^{\Omega \Omega} = A l^{- \alpha}$ where $\alpha = n + 3$ and $A$ is 
the amplitude which depends on the source redshift. Motivated by this fact, we assume power law forms of 
$C_{l}^{\Omega \Omega}$ characterized by an amplitude $A$, power $\alpha$ and a cutoff in $l$, 
denoted by $l_{max}$. Given an $l_{max}$ and $\alpha$, we determine the value of $A$ which will produce a 
detectable effect on lensed $C_{l}^{BB}$.
We denote the lensing contribution of $C_{l}^{\Omega \Omega}$
to $C_{l}^{BB}$ by $\delta \tilde C_{l}^{BB}$. To obtain this threshold we compare 
$\delta \tilde C_{l}^{BB}$ with the cosmic variance.
For a given $\alpha$, we want to know the 
value of amplitude $A$ for which maxima of $\delta \tilde C_{l}^{BB}$ reaches a particular 
value. This particular value is chosen to be three times the value of the cosmic variance 
of lensed $C_{l}^{BB}$ due to $C_{l}^{\phi \phi}$ at the multipole
where the maxima occur. 
This is an idealistic criteria which assumes zero noise experiment limited only by cosmic variance. 

\par Once the constrained form of $C_{l}^{\Omega \Omega}$ is known, we use it to obtain the 
constrained form of $P_{H}(k)$. 
\Eq{eq_clomegaomega} shows that $C_{l}^{\Omega \Omega}$ is convolution of $P_{H}(k)$ 
and $|T_{l}^{H}(k)|^{2}$. A given a set of cosmological parameters completely determines $T_{l}^{H}(k)$. 
To obtain $P_{H}(k)$ for given $C_{l}^{\Omega \Omega}$ and $T_{l}^{H}(k)$ we use the 
Richardson-Lucy (RL) deconvolution algorithm \cite{RICHARDSON:72, Lucy:1974yx}
This method has been used
in the literature to deconvolve primordial power spectrum of scalar 
perturbations using WMAP and Planck data of CMB temperature anisotropies
\cite{PhysRevD.70.043523, Hazra:2013xva, Hazra:2014jwa, 1475-7516-2009-07-011}. 
To apply this method we write \Eq{eq_clomegaomega} in discrete form
\begin{eqnarray}\label{eq_discrete_clomega}
 C_{l}^{\Omega \Omega} = \sum_{i} G(l,k_{i}) P_{H}(k_{i}),
\end{eqnarray}
where 
\begin{eqnarray}
 G(l,k_{i}) = \frac{4 \pi^2}{l^{2}(l+1)} \frac{(l+2)!}{(l-2)!} \Delta k_{i} k_{i}^{2} |T^{H}_{l}(k_{i})|^{2}.
\end{eqnarray}
Given $C_{l}^{\Omega \Omega}$, $G(l,k_{i})$ and the initial guess for 
$P_{H}(k)$, RL method iteratively solves for the power spectrum using the 
following relation 
\begin{eqnarray}
 P_{H}^{r+1}(k_i) = P_{H}^{r}(k_i)
 + P_{H}^{r}(k_i) \sum_{l} G(l, k_i) \frac{C^{\Omega \Omega}_{l}-C^{(r)}_{l}}{C^{(r)}_{l}}
 \end{eqnarray}
at each $k_i$. Here $P_{H}^{r}(k_i)$ is the power spectrum obtained after $r^{th}$ 
iteration. $C^{r}_{l}$ is the $C_{l}^{\Omega \Omega}$ recovered using 
\Eq{eq_discrete_clomega} for $r^{th}$ iterate of the spectrum, $P^{r}_{H}(k)$
\begin{eqnarray}
 C_{l}^{r} = \sum_{i} G(l,k_{i}) P^{r}_{H}(k_{i}).
\end{eqnarray}

We monitor the sum of square of relative error between recovered $C^{r}_{l}$ and input
$C_l$ to decide when to stop the iterations. The iterations are carried out until the quantity 
\begin{eqnarray}
 \sigma^{2}  = \sum_{l} \Big (\frac{C^{\Omega \Omega}_{l} - C^{r}_{l}}{C_{l}^{\Omega \Omega}}\Big )^{2}
\end{eqnarray}
reaches a particular predetermined value. 
This controls the accuracy of recovered power spectrum. 
For our analysis we have taken the value of $\sigma^{2}$ such that the discrepancy of the recovered 
$P_{H}(k)$ will translate to negligible difference in the value of lensed $C_{l}^{BB}$. This 
discrepancy is set well below the cosmic variance of the lensed $C_{l}^{BB}$.
We have tested our algorithm by implementing it on the $C_{l}^{\Omega \Omega}$ to recover $P_{H}(k)$ 
that is known beforehand. Our implementation of RL algorithm could recover $P_{H}(k)$ within 
above mentioned accuracy. The recovered $P_{H}(k)$ has wiggles peculiar to RL algorithm. We smooth 
out the wiggles in recovered $P_{H}(k)$. This $P_{H}(k)$ is then used to obtain the 
$\Omega_{GW}(k)$ using \Eq{eq_Omega_GW}. 

\section{Results}
\begin{figure}
    \subfigure[]{\label{fig_A_alpha_a}  
\includegraphics[height=6cm,width=8cm,angle=0]{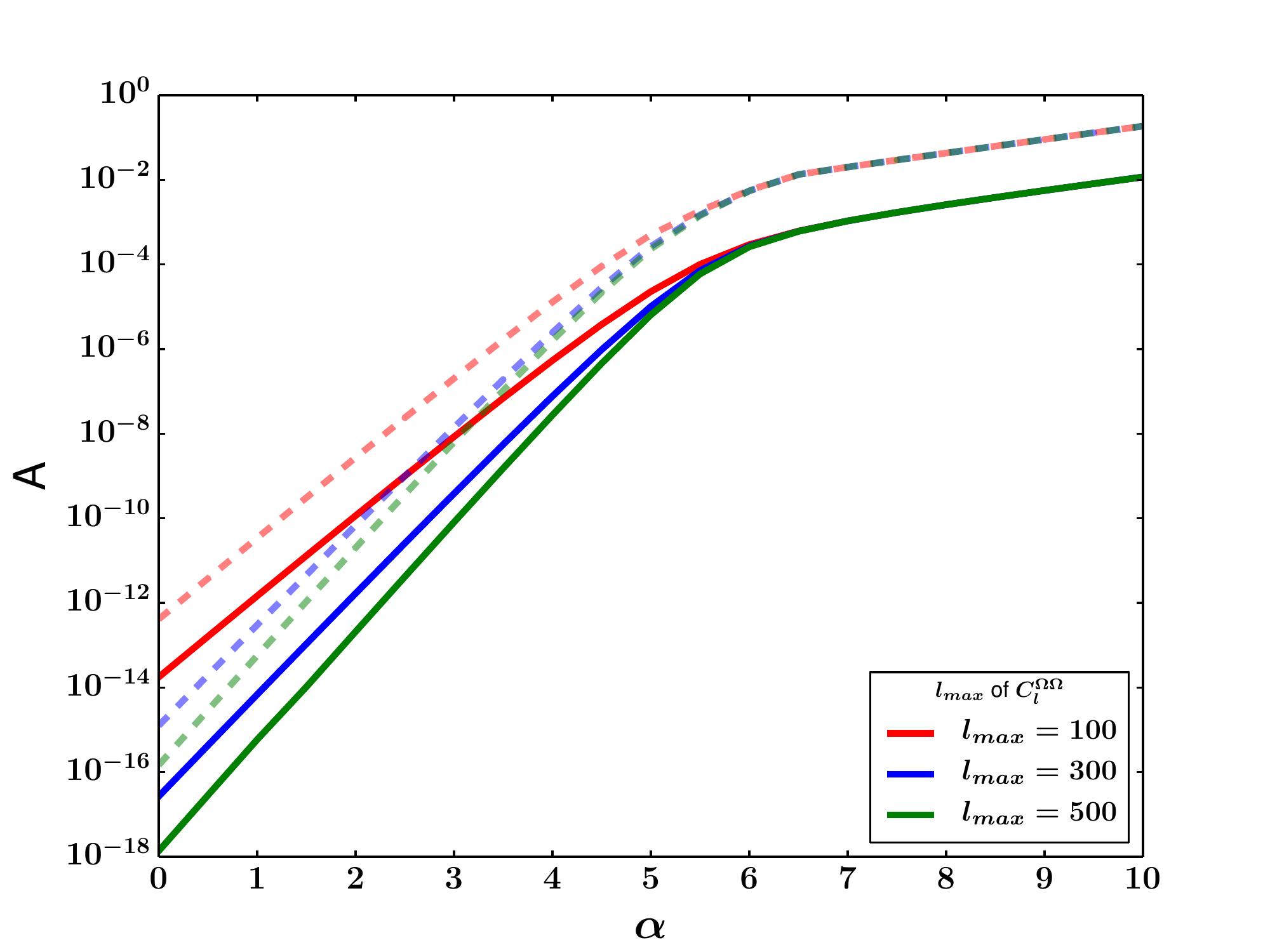}}
    \subfigure[]{\label{fig_A_alpha_b}  
\includegraphics[height=6cm,width=8cm,angle=0]{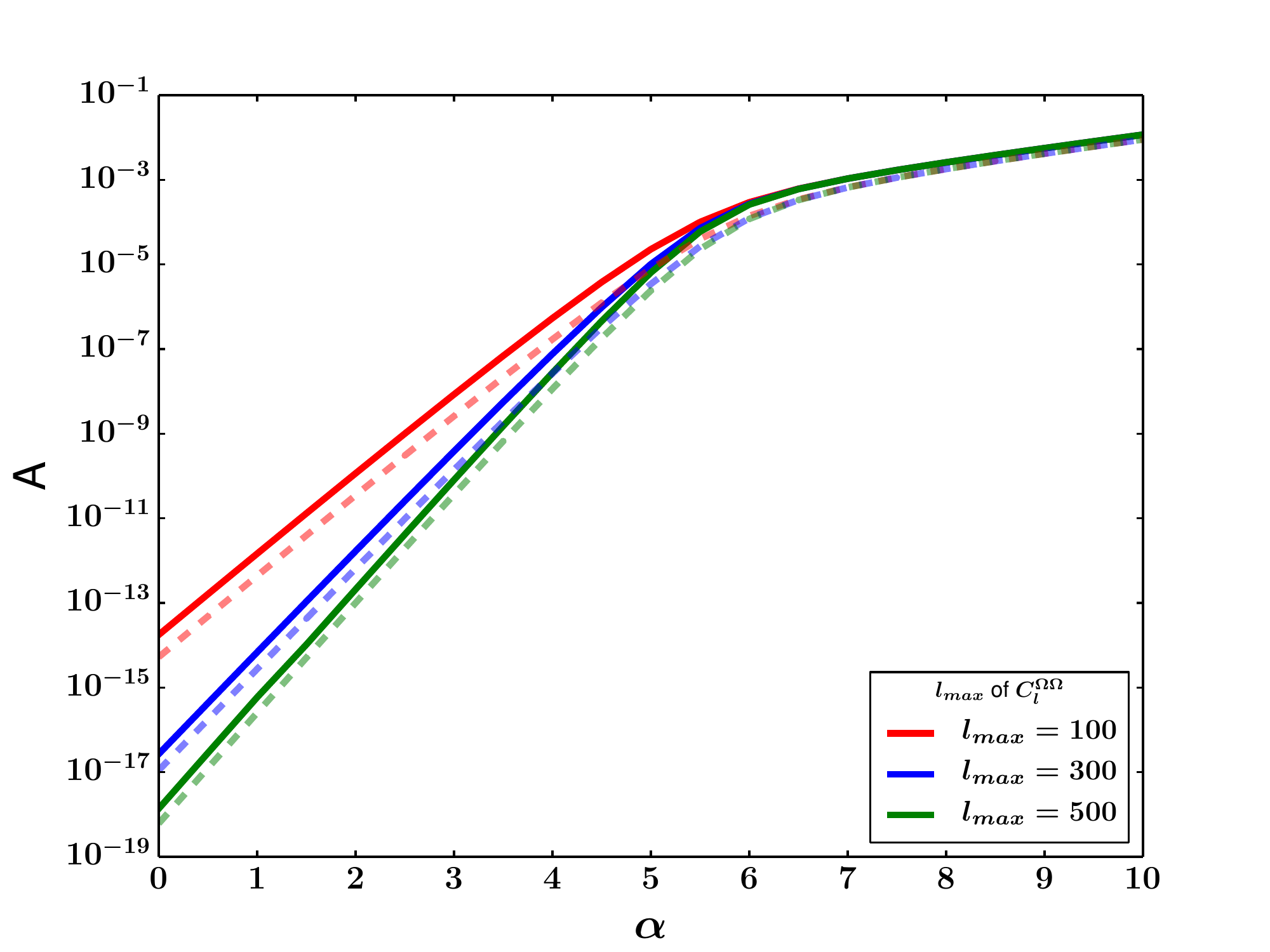}}
		\captionsetup{singlelinecheck=off,justification=raggedright}
	 \caption{For the power-law form of $C_l^{\Omega\Omega} = Al^{-\alpha}$ as discussed in Sec. \ref{methods}, 
	 we obtain the bounds on parameters $A$ for different values of $\alpha$ and $l_{max}$.
	 (a) Continuous curves are for bounds using $C_{l}^{BB}$ while dashed curves are bounds 
using $C_{l}^{TT}$. Bounds obtained using $C_{l}^{BB}$ are stronger than those of $C_{l}^{TT}$. For 
$\alpha > 6$, bounds on $A$ are insensitive to $l_{max}$.
	 (b) Continuous curves are for bounds using $C_{l}^{BB}$ with rotation and dashed 
curves are for bounds using $C_{l}^{BB}$ without rotation.}
	  \label{fig_A_alpha}
\end{figure}

In Fig. \ref{fig_A_alpha} we give bounds on $A$ for values of 
$\alpha$ ranging from $0$ to $10$. Results for different $l_{max}$ cutoff 
are given. Within the power law approximation considered here $\alpha = 6$ corresponds 
to $P_{H}(k) = k^{-3}$, which is scale invariant power spectrum. Hence $\alpha 
> 6$ corresponds to red $P_{H}(k)$ whereas $\alpha < 6$ corresponds to blue $P_{H}(k)$. 
As a consequence bounds on $A$ are expected to be less sensitive to value of $l_{max}$ 
for $\alpha > 6$. This is evident from the Fig. \ref{fig_A_alpha}. For $\alpha > 6$, all the curves corresponding 
to different $l_{max}$ give same bound on $A$. We carry out the same exercise with lensing of 
$C_{l}^{TT}$ and obtain the bounds on $A$, also shown in Fig. \ref{fig_A_alpha_a}. The bounds on 
$A$ obtained using $C_{l}^{TT}$ are weaker roughly by one order of magnitude compared to the 
bounds from $C_{l}^{BB}$. In Fig. \ref{fig_A_alpha_b} we depict the bounds on $A$ with and without 
the rotation term obtained using $C_{l}^{BB}$. At higher values of $\alpha$ the value of $A$ 
becomes less sensitive to the rotation term. $C_{l}^{\Omega \Omega}$ with lower values of $\alpha$ 
have  large power at low $l$ compared to those with higher values of $\alpha$. This leads to 
cancellation effect of rotation term being more effective for low $\alpha$ values compared to 
higher $\alpha$ values.
Bounds obtained using $C_{l}^{TT}$ are not affected 
by rotation because rotation of polarization do not affect $C_{l}^{TT}$.

Given $\alpha$, and corresponding constrained $A$, we obtain allowed forms of $C_{l}^{\Omega \Omega}$. 
We use the RL algorithm to reconstruct the constrained $P_{H}(k)$. 
Given an $l_{max}$ and source 
redshift $z_{s}$, $P_{H}(k)$ can be constrained only up to $k_{max} = l_{max}/(\eta_s - \eta_0)$. 
Hence, larger the source redshift smaller is the $k_{max}$ up to which we can constrain $P_{H}(k)$. 

\par For any physical model, we expect a natural cut off in wavenumber ($k_{max}$) up to which $P_H(k)$ is 
non-zero. For $\alpha > 6$ which corresponds to red spectra, $P_H(k)$ decreases with $k$ 
and for blue spectra with $\alpha < 6$, $P_H(k)$ increases with $k$.
Power spectrum of tensor perturbations from inflation as well as by second order 
effects in density perturbations both are red at the $k$ range we are interested in 
\cite{devdeep_s_cosmic_shear}. The power spectrum of \cite{Julian_Adamek_and_Ruth_Durrer} for 
tensor perturbations from second order effects generated at various redshifts is also red in nature. 
Blue spectrum is unlikely to be produced by such physical mechanisms at the 
$k$ range we are interested. So, we restrict our estimation of $\Omega_{GW}$ for red spectra, which correspond to $\alpha   
\geq 6$. This also ensures that we do not need to make any model dependent choice of $k_{max}$. We also note that 
the individual spectrum of tensor perturbations 
 predicted in \cite{devdeep_s_cosmic_shear, Julian_Adamek_and_Ruth_Durrer} are not strong enough to 
contribute to detectable levels of lensed B-modes.

\begin{figure}[H]
    \subfigure[Source redshift $z_s = 1$]{\label{fig_ogw_a}  
\includegraphics[height=6cm,width=8cm,angle=0]{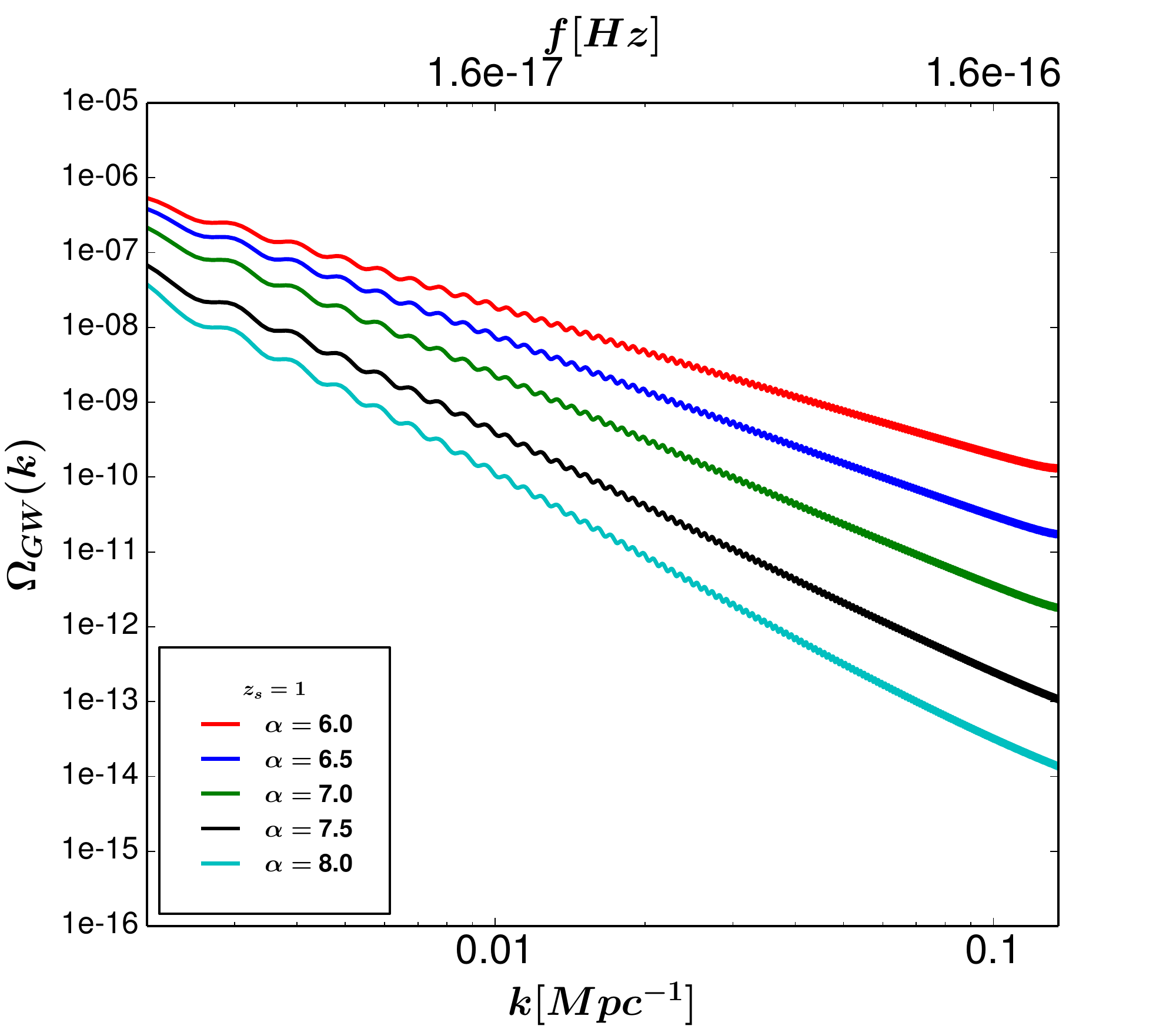}}
    \subfigure[Source redshift $z_s = 10$]{\label{fig_ogw_b}  
\includegraphics[height=6cm,width=8cm,angle=0]{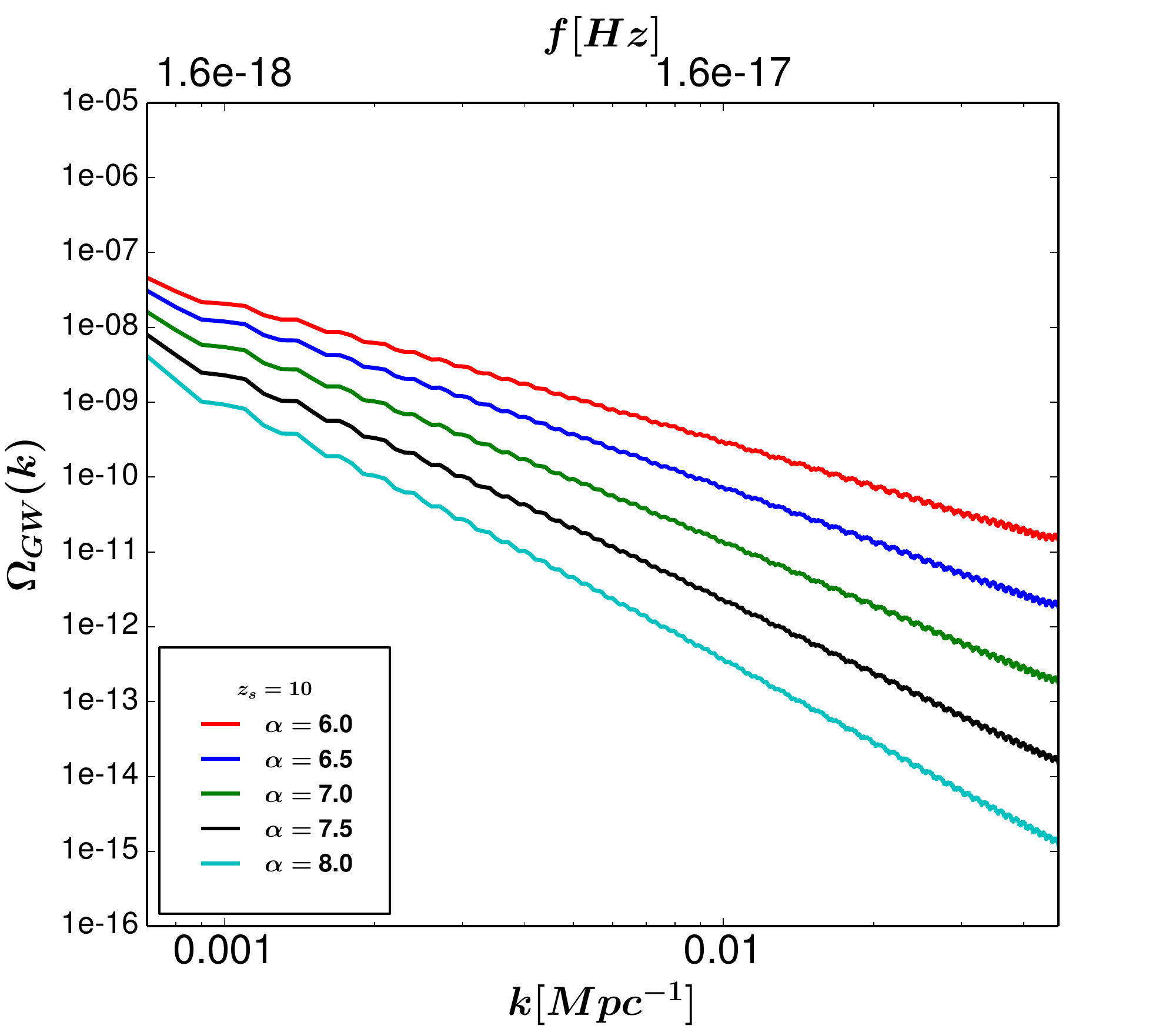}}
		\captionsetup{singlelinecheck=off,justification=raggedright}
	 \caption{Constrained $\Omega_{GW}(k)$ for two source redshifts, $z_s = 1$ and $z_s = 10$. 
	 $\Omega_{GW}$ corresponding to red $P_{H}(k)$ obtained for different values of $\alpha$ 
and $l_{max} = 500$ are shown. For given $l_{max}$ and $z_s$, $\Omega_{GW}$ is constrained up to 
$k_{max} = l_{max}/(\eta_0 - \eta_s)$. The results are given for the range of $k$ over which our 
recovery is faithful.}
	  \label{fig_ogw}
\end{figure}

\par We use \Eq{eq_Omega_GW} to get $\Omega_{GW}(k)$ corresponding to reconstructed $P_{H}(k)$.  
Fig. \ref{fig_ogw} represent $\Omega_{GW}(k)$ for two source redshifts obtained using $C_{l}^{BB}$. 
We have taken the example of $l_{max} = 500$ to elucidate our method.
Curves shown in Fig. \ref{fig_ogw} are obtained 
using the running bin average of actual $\Omega_{GW}$ to reduce the wiggles which would otherwise 
be present due to the oscillatory behavior of the term $\frac{dT_{H}(x)}{dx}$ in \Eq{eq_Omega_GW}.
In Fig. \ref{fig_ogw} 
we see that for given $\alpha$, $\Omega_{GW}$ for redshift $z_s = 10$ is smaller than that of 
redshift $z_s = 1$. This is expected because to obtain a given amount of $C_{l}^{\Omega \Omega}$ 
one needs small power at high redshift than that at lower redshift. 

\section{Conclusion}
Previous work argued that B-mode generated due to photon deflection are largely canceled by the 
rotation induced by tensor perturbations. Here we have demonstrated that this result is not generic 
and depend on the specifics of $C_{l}^{\Omega \Omega}$. The contribution of the rotation of 
polarization depends on the relative contribution of $C_{l}^{\psi \psi}$ and $C_{l}^{\psi \Omega}$ 
terms.  Rotation term may contribute to lensing through subtraction or addition depending on the nature of 
curl deflection potential $C_{l}^{\Omega \Omega}$.
Specifically, we note that the rotation term is most efficient at reducing power transfer from 
E-modes to B-modes when the power in $C_{l}^{\Omega \Omega}$ is concentrated at low $l$. Whereas, 
presence of more power at high $l$ in $C^{\Omega\Omega}_{l}$ decreases this efficiency (as depicted 
in Fig. \ref{fig1}).

The weak lensing of the CMB due to SGWB provides us a window to constrain SGWB of cosmological 
origin. In this work, we have exploited this effect to derive upper bounds on the energy density $\Omega_{GW}$ of the SGWB. 
To derive these constraints, we do not assume any particular model for the origin of SGWB, except 
that we present our constraints only for red spectra $P_{H}(k)$.
We constrain the form of $\Omega_{GW}$ using idealistic constraints on 
$C_{l}^{\Omega \Omega}$.  We first constrain the power law forms of $C_{l}^{\Omega \Omega}$ 
and translate it into upper bounds on $\Omega_{GW}$ sourced at a given redshift. In this paper, we 
present the model independent upper bound on $\Omega_{GW}$ 
spectrum which can lead to a particular observable imprint in CMB.
Any model predicting $\Omega_{GW}(k)$ more than the ones depicted in Fig. \ref{fig_ogw} over the 
range 
of $k$ will be able to cast an observable signature on CMB $B$ mode polarization through lensing. 

\section{Acknowledgements}
S.S. acknowledges University Grants Commission (UGC), India for providing the financial support  as Senior Research Fellow. S.M. thanks Council of Scientific \& Industrial Research (CSIR), India for financial support as Senior Research Fellow. The present work is carried out using the High Performance Computing facility at IUCAA.


\label{Bibliography}


\bibliographystyle{JHEP}
\bibliography{reference_inspirehep_bibtex}
\end{document}